\documentclass[twocolumn,showpacs,showkeys,preprintnumbers,amsmath,amssymb,prb]{revtex4-1}

\usepackage{graphicx}
\usepackage{bm}
\usepackage{color}
\usepackage{epstopdf}
\usepackage[latin2]{inputenc} 

\usepackage[urlcolor=blue,colorlinks=true,citecolor=blue,linkcolor=blue,pdfstartview={FitH},bookmarks=false]{hyperref}

\begin{document}
\title{Unconventional superconductivity in iron-base superconductors in a three-band model}
\author{Dawid Crivelli}
\email[E-mail: ]{dcrivelli@us.edu.pl}
\author{Andrzej Ptok}
\email[E-mail: ]{aptok@mmj.pl}
\affiliation{Institute of Physics, University of Silesia, 40-007 Katowice, Poland}

\begin{abstract}
Iron-base superconductors exhibits features of systems where the Fulde-Ferrel-Larkin-Ovchinnikov (FFLO) phase, a superconducting state with non-zero total momentum of Cooper pairs, is actively sought. Experimental and theoretical evidence points strongly to the FFLO phase in these materials above the Pauli limit. In this article we discuss the ground state of iron-base superconductors near the critical magnetic field and the full $h-T$ phase diagram for pnictides in case of intra-band pairing, in a three-band model with $s_{\pm}$ symmetry.
\end{abstract}


\maketitle

\section{Introduction}

In '60s of the XX century, two independent groups, Fulde-Ferrell (FF)~\cite{FF} and Larkin-Ovchinnikov (LO)~\cite{LO}, proposed a superconducting phase with oscillating order parameter (OP) in real space. This phase, nowadays called the Fulde-Ferrell-Larkin-Ovchinnikov (FFLO) phase, is more stable than the BCS phase in low temperature and hight magnetic field regime. FF proposed a superconducting phase with one momentum ${\bm q}$ possible for Cooper pairs, whereas LO assumed the possibility of two opposite momenta $\pm {\bm q}$ -- in this case the OP in real space is proportional to $\exp ( i {\bm q} \cdot {\bm r} )$ or $\cos ( {\bm q} \cdot {\bm r} )$ respectively. A non-zero total momentum of Cooper pairs bears as a consequence the change of sign of the order parameter (OP) in real space and breaks the spatial symmetry of the system (this is true not only in systems with translation symmetry, but also when rotational symmetry is present~\cite{yanase.09,iskin.williams.08,ptok.12}).

The FFLO phase can be expected in materials with relatively high Maki parameter $\alpha \sim H_{c2}^{orb} / H_{c2}^{P}$, when the orbital critical magnetic field $H_{c2}^{orb}$ is greater than the paramagnetic critical field $H_{c2}^{P}$. Therefore a good class of candidate to find the FFLO are heavy fermions materials
(such as $CeCoIn_{5}$)~\cite{capan.bianchi.04,miclea.nicklas.06, bianchi.movshovich.03,martin.agosts.05,correa.murphy.07, kakuyanagi.saitoh.05,matsuda.shimahara.07}, 
organic superconductors~\cite{lortz.wang.07} and quantum gases~\cite{casalbuoni.nardulli.04}. The FFLO phase can exist also in inhomogeneous systems in presence of impurities~\cite{wang.hu.06,wang.hu.07,ptok.10} or spin density waves~\cite{ptok.maska.11}. Moreover these inhomogeneities can increase the tendency system to create the FFLO phase and stabilize it in a lower magnetic field.~\cite{ptok.10,mierzejewski.ptok.10} The FFLO phase can be also stabilized by pair hopping interaction~\cite{ptok.mierzejewski.08,ptok.maska.09} or in system with nonstandard quasiparticles with spin-dependent mass.~\cite{kaczmarczyk.jedrak.10,kaczmarczyk.spalek.10,maska.mierzejewski.10}

\begin{figure}[!b]
\includegraphics{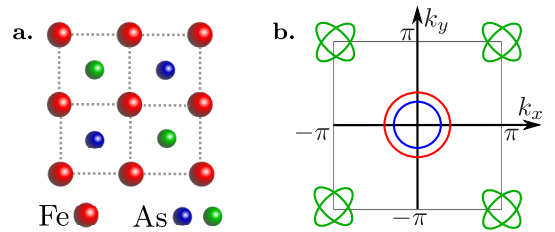}
\caption{(Color on-line) (Panel a) $\mathtt{FeAs}$ layer in iron-base superconductors. $\mathtt{Fe}$ (red dot) and $\mathtt{As}$ (blue and green dots) ions form a quadratic lattice. $\mathtt{As}$ ions are placed above (blue) or under (green) the centers of the squares formed by $\mathtt{Fe}$. (Panel b) True Fermi surface in first Brillouin zone, for two $\mathtt{Fe}$ ions per unit cell.}
\label{fig.feas}
\end{figure}

Other good candidates to find the FFLO phase are iron-based superconductors (IBSC)~\cite{kamihara.watanabe.08,gurevich.10,gurecich.11,ptok.crivelli.13,ptok.14} -- the characteristic feature of these chemical compounds are iron-arsenide layers (Fig. \ref{fig.feas}.a), which imply multi-band properties such as the characteristic Fermi surface (with hole- and electron-like Fermi pockets around the $(0,0)$ and $(\pi,\pi)$ point respectively, illustrated in Fig. \ref{fig.feas}.b).~\cite{singh.du.08,ding.richard.08,kondo.santander.08} IBSC are materials with high Maki parameter and anisotropic upper magnetic fields.~\cite{fuchs.drechsler.08,terashima.kimata.09,kurita.kitagawa.11,
cho.kim.11,zhang.jiao.11,khim.lee.11,liu.tanatar.13,burer.hardy.13}
Experimentally a phase transition inside the superconducting state has been observed, which can be evidence about the phase transition from convectional superconductivity to the FFLO phase.~\cite{zocco.grube.13} These results are agreement with theoretical expectations.~\cite{ptok.14,mizushima.takahashi.13,takahaski.mizushima.14}

In this paper we analyze IBSC (pnictides) using the three band model proposed by M. Daghofer {\it et al}.~\cite{daghofer.nicholson.10,daghofer.nicholson.12} In section \ref{sec.theory} we describe details of theoretical calculation, in section \ref{sec.numeric} we show and discuss numerical results. We summarize the results in section \ref{sec.summary}. Parameters for the model are listed in Appendix \ref{app.threeband}.

\section{Theoretical part}
\label{sec.theory}

The general Hamiltonian for the multi-orbital system can be written as $H = H_{0} + H_{I}$. The non-interacting part $H_{0}$ is given by:
\begin{eqnarray}
H_{0} &=& \sum_{{\bm k}\sigma,\alpha\beta} \left( T_{\bm k}^{\alpha\beta} - ( \mu + \sigma h ) \delta_{\alpha,\beta} \right) c_{{\bm k}\alpha\sigma}^{\dagger} c_{{\bm k}\beta\sigma}
\end{eqnarray}
where $c_{{\bm k}\alpha\sigma}^{\dagger} (c_{{\bm k}\alpha\sigma}) $ is the creation (annihilation) operator for a spin $\sigma$ electron of momentum ${\bm k}$ in the orbital $\alpha$. Hopping matrix elements $T_{\bm k}^{\alpha\beta}$ are given by the effective tight-binding model of the two dimensional $\mathtt{FeAs}$ planes in the given model (see Appendix \ref{app.threeband}). Integer $\alpha$ and $\beta$ label the orbitals. Band structure of the $\mathtt{FeAs}$ system can be reconstructed by diagonalization of the Hamiltonian $H_{0}$:
\begin{eqnarray}
H'_{0} &=& \sum_{{\bm k}\varepsilon\sigma} E_{{\bm k}\varepsilon\sigma} d_{{\bm k}\varepsilon\sigma}^{\dagger} d_{{\bm k}\varepsilon\sigma} .
\end{eqnarray}
$\mu$ is the chemical potential, changing the average number of particles in the system $n = \frac{1}{N} \sum_{{\bm k}\alpha\sigma} c_{{\bm k}\alpha\sigma}^{\dagger} c_{{\bm k}\alpha\sigma} = \frac{1}{N} \sum_{{\bm k}\alpha\sigma} d_{{\bm k}\alpha\sigma}^{\dagger} d_{{\bm k}\alpha\sigma} $, where $N$ is the number of lattice site. $h$ is the external magnetic field parallel to lattice. $\varepsilon$ labels the bands.

We introduce a superconducting pairing between {\it quasi}-particles in bands $\varepsilon$. In absence of interband pairing or when it is weak,~\cite{mazin.10} we can effectively describe superconductivity in the FFLO phase by the Hamiltonian:
\begin{eqnarray}
H'_{SC} &=& \sum_{\varepsilon{\bm k}} \left( \Delta_{\varepsilon{\bm k}} d_{\varepsilon{\bm k}\uparrow}^{\dagger} d_{\varepsilon,-{\bm k}+{\bm q}_{\varepsilon} \downarrow}^{\dagger} + H.c. \right ) ,
\end{eqnarray}
where $\Delta_{\varepsilon{\bm k}} = \Delta_{\varepsilon} \eta ( {\bm k} )$ is the amplitude of the OP for Cooper pairs with total momentum ${\bm q}_{\varepsilon}$.
The structure factor is given by  $\eta ( {\bm k} ) = 4 \cos( k_{x} ) \cos ( k_{y} )$ for $s_{\pm}$-{\it wave} symmetry of the OP.~\cite{ptok.crivelli.13} As we see, in case intra-band pairing we have formally an $n$-band system described by the total Hamiltonian $H = H'_{0} + H'_{SC}$, with $n$ independent bands $\varepsilon$. Using the Bogoliubov transformation we can find a final fermions basis $\Gamma_{\varepsilon{\bm k}} = ( \gamma_{\varepsilon{\bm k}\uparrow} , \gamma_{\varepsilon,-{\bm k}\downarrow} )^{T}$, describing the quasi-particle excitation in the superconducting state:
\begin{eqnarray}
H = \sum_{\varepsilon{\bm k}\tau} \bar{E}_{\varepsilon{\bm k}\tau} \gamma_{\varepsilon{\bm k}\tau}^{\dagger} \gamma_{\varepsilon{\bm k}\tau} + const.
\end{eqnarray}
with
\begin{eqnarray}
\bar{E}_{\varepsilon{\bm k}\tau} &=& \frac{ E_{\varepsilon{\bm k}\uparrow} - E_{\varepsilon,-{\bm k}+{\bm q}\downarrow} }{2} \\
\nonumber &+& \tau \sqrt{ \left( \frac{ E_{\varepsilon{\bm k}\uparrow} + E_{\varepsilon,-{\bm k}+{\bm q}\downarrow} }{2} \right)^{2} + | \Delta_{\varepsilon{\bm k}} |^{2} } 
\end{eqnarray}
where $\tau = \pm$. Total free energy is given by $\Omega = \sum_{\varepsilon} \Omega_{\varepsilon}$, where:
\begin{eqnarray}
\Omega_{\varepsilon} &=& -k_{B} T \sum_{{\bm k}\tau} \ln \left( 1 + \exp ( - \beta \bar{E}_{\varepsilon{\bm k}\tau} ) \right) \\
\nonumber &+& \sum_{\bm k} \left( E_{\varepsilon{\bm k}\downarrow} - \frac{ | \Delta_{\varepsilon{\bm k}} |^{2} }{ V_{\varepsilon} } \right) .
\end{eqnarray}
is the free energy in band $\varepsilon$ in the presence effective interaction intensity $V_{\varepsilon}$. 
The ground state for fixed $h$ and $T$ can be found by minimizing the free energy w.r.t. the OPs.

\section{Numerical results}
\label{sec.numeric}

Numerical calculations were carried out for a square lattice $N_{X} \times N_{Y} = 2000 \times 2000$ with periodic boundary conditions. First, the effective pairing intra-band potential $V_{\varepsilon}$ has been determined for every band, in case of $s_{\pm}$ symmetry of the order parameter -- to find its value we seek the disappearance of the superconducting BCS phase in each band at the same critical magnetic field $h_{C}^{BCS} = 0.005 [eV]$ (and temperature $k_{B} T = 10^{-5} [eV]$). Secondly, we determine the $h-T$ phase diagram for those fixed values.

\begin{figure}[!b]
\includegraphics{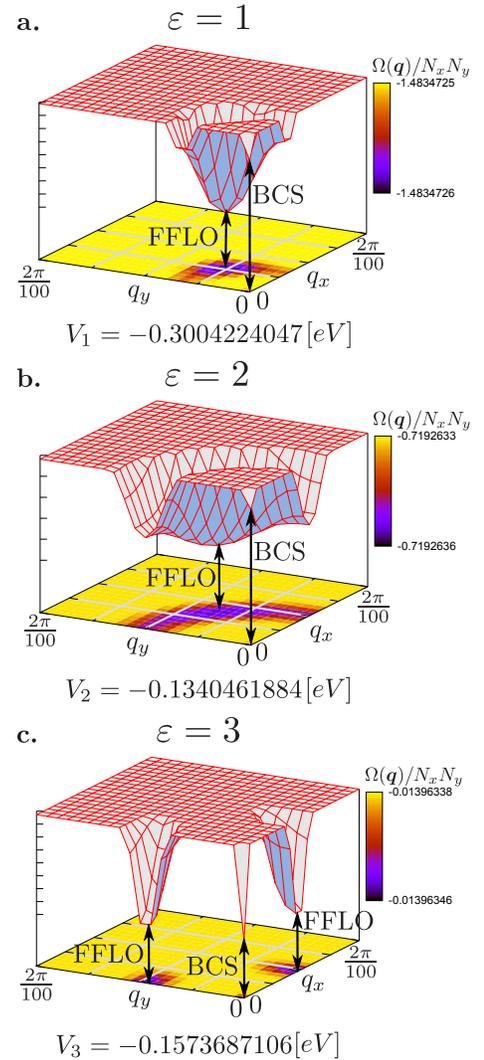}
\caption{(Color on-line) The free energy per site $\Omega_{\varepsilon} ({\bm q})/N_{x}N_{y}$ for $s_{\pm}$ symmetry, for different values of the Cooper pair momentum ${\bm q}$, showing the location of the minima and indicating the existence of different phases. Results for $h \simeq 0.005 [eV] = h_{C}^{BCS}$ and temperature $k_{B} T = 10^{-5} [eV]$.}
\label{fig.gs}
\end{figure}

\paragraph*{Ground state at the BCS critical magnetic field.} To determine the $h-T$ phase diagram with $V_{\varepsilon}$ fixed, we vary the total momentum of Cooper pairs ${\bm q}$ to find the ground state. Results for magnetic field $h \simeq h_{C}^{BCS}$ and temperature $k_{B} T = 10^{-5} [eV]$ are shown in Fig.~\ref{fig.gs}. As we see for every band and ${\bm q} = 0$, there exists a local minimum of the free energy $\Omega_{\varepsilon} ( {\bm q} )$ corresponding to the BCS phase. However we find the true ground state by the global minimum, which is attained for ${\bm q} \neq 0$. For the first two bands ($\varepsilon = 1,2$ -- panels a and b respectively) the ground state can be found for four equivalent total momenta ${\bm q}_{1,2}$ in directions $[1,\pm1]$. In the third band ($\varepsilon =3$ -- panels c) the global minimum also exists at non-zero total momentum of Cooper pairs, but in direction $[0,1]$ or $[1,0]$. This result is in agreement with other theoretical results for pnictides in a minimal two-band model~\cite{ptok.crivelli.13,ptok.14} and one-band heavy fermions systems,~\cite{matsuda.shimahara.07,ptok.maska.11,mierzejewski.ptok.10,ptok.maska.09} where the FFLO phase exhibits precisely this direction of the momentum.

\begin{figure}[!b]
\includegraphics{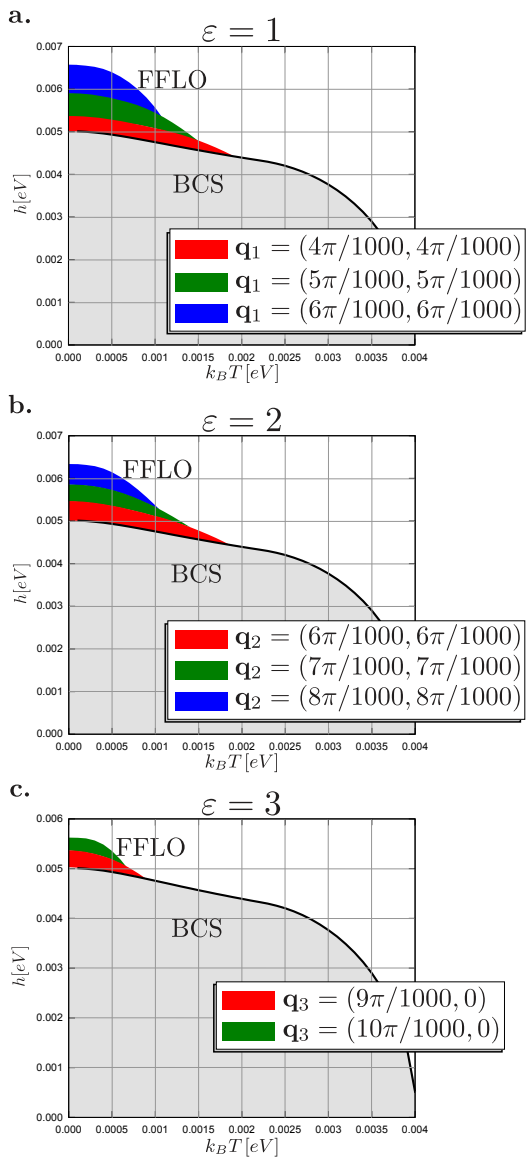}
\caption{(Color on-line) $h-T$ phase diagram for given effective pairing potential $V_{\varepsilon}$. Grey area shows region of existing BCS phase. 
Lines mark the phase transitions. Colors (red, green and blue) mark the regions of the FFLO phase with different values of the total momentum of Cooper pairs ${\bf q}_{\varepsilon}$.
}
\label{fig.df}
\end{figure}

\paragraph*{Phase diagram $h-T$.} For fixed values of the effective pairing potential $V_{\varepsilon}$, we find the $h-T$ phase diagram for each band, as shown in Fig. \ref{fig.df}. The region of the BCS phase on the phase diagram has a typical form. Above the critical magnetic field for BCS phase and at low temperatures, the FFLO phase can form (cyan region in Fig. \ref{fig.df}). In the first two bands, the critical magnetic field of the FFLO phase is bigger than in the third band (Fig \ref{fig.df}) -- for the chosen values of $V_{\varepsilon}$ this difference is approximately equal to $\frac{1}{5} h_{C}^{BCS}$.

\paragraph*{Total momentum of Cooper pairs.} 
Minimization of the free energy gives the total momentum of Cooper pairs (shown in Fig. \ref{fig.gs} in magnetic field near $h_{C}^{BCS}$). Its value $| {\bf q}_{\varepsilon} |$ is higher for the first band than for the second and third bands. However an increase in the magnetic field raises the total momentum magnitude (red, green and blue areas in Fig. \ref{fig.df}), as in the IBSC two-band model.~\cite{ptok.14}

In every band the critical magnetic fields of the phase transition from the FFLO phase to normal state $h_{C}^{FFLO} (T)$ are different. Consequence of this are the observed multiple transitions inside the FFLO area of the phase diagram, associated with changes in the modules of total momentum of Cooper pairs $| {\bm q}_{\varepsilon} |$. Moreover this leads to amplitude modulation of the order parameter in real space, in agreement with the results in in the two-band case.~\cite{ptok.14,mizushima.takahashi.13,takahaski.mizushima.14} To observe this feature would be an experimental check of the existence of the FFLO phase in these materials,~\cite{zocco.grube.13} since we expect more than one phase transition to exist, associated with disappearance of the FFLO phase in selected bands when increasing the external magnetic field $h$.

\section{Summary}
\label{sec.summary}

Using the three-band model proposed by Daghofer {\it et al.}~\cite{daghofer.nicholson.10,daghofer.nicholson.12} we make a case for the FFLO phase in iron-base superconductors in presence of intra-band pairing with $s_{\pm}${\it -wave} symmetry.
As in previous theoretical works,~\cite{ptok.crivelli.13,ptok.14} we show that the ground state of pnictides, above the critical magnetic field of BCS phase and in low temperature, is an unconventional superconductor of the FFLO type. The full phase diagram has been obtained on lattices of thermodynamically relevant sizes, marking the typical area of the BCS phase and how the FFLO can be found beyond its borders, in regimes detrimental to the existence of BCS superconductivity. Consequence of this is the amplitude modulation of the order parameter in real space and multiple phase transitions, in agreement with the literature.~\cite{ptok.14,mizushima.takahashi.13,takahaski.mizushima.14}

\begin{acknowledgments}
D.C. acknowledges support by the FORSZT PhD fellowship.
\end{acknowledgments}

\appendix

\section{Three-orbital model Daghofer {\it et al.}}
\label{app.threeband}


This model of IBSC was proposed by Daghofer {\it et al.} in Ref. [\onlinecite{daghofer.nicholson.10}] and improved in Ref. [\onlinecite{daghofer.nicholson.12}]. Beyond $d_{xz}$ and $d_{yz}$ orbitals the model also accounts for $d_{xy}$ orbital:
\begin{eqnarray}
T_{\bm k}^{11} &=& 2 t_{2} \cos k_{x} + 2 t_{1} \cos k_{y} + 4 t_{3} \cos k_{x} \cos k_{y} \\
\nonumber &+& 2 t_{11} ( \cos ( 2 k_{x} ) - \cos ( 2 k_{y} ) ) + 4 t_{12} \cos ( 2 k_{x} ) \cos ( 2 k_{y} ) ,
\end{eqnarray}
\begin{eqnarray}
T_{\bm k}^{22} &=& 2 t_{1} \cos k_{x} + 2 t_{2} \cos k_{y} + 4 t_{3} \cos k_{x} \cos k_{y} \\
\nonumber &-& 2 t_{11} ( \cos ( 2 k_{x} ) - \cos ( 2 k_{y} ) ) + 4 t_{12} \cos ( 2 k_{x} ) \cos ( 2 k_{y} ) ,
\end{eqnarray}
\begin{eqnarray}
\nonumber T_{\bm k}^{33} &=& \epsilon_{0} + 2 t_{5} ( \cos k_{x} + \cos k_{y} ) + 4 t_{6} \cos k_{x} \cos k_{y} \\
&+& 2 t_{9} ( \cos ( 2 k_{x} ) + \cos ( 2 k_{y} ) ) \\
\nonumber &+& 4 t_{10} ( \cos ( 2 k_{x} ) \cos k _{y} + \cos k_{x} \cos ( 2 k_{y} ) ) ,
\end{eqnarray}
\begin{eqnarray}
T_{\bm k}^{12} &=& T_{\bm k}^{21} = 4 t_{4} \sin k_{x} \sin k_{y} ,
\end{eqnarray}
\begin{eqnarray}
T_{\bm k}^{13} &=& \bar{T}_{\bm k}^{31} = 2 i t_{7} \sin k_{x} + 4 i t_{8} \sin k_{x} \cos k_{y} ,
\end{eqnarray}
\begin{eqnarray}
T_{\bm k}^{23} &=& \bar{T}_{\bm k}^{32} = 2 i t_{7} \sin k_{y} + 4 i t_{8} \sin k_{y} \cos k_{x} .
\end{eqnarray}
In Ref. [\onlinecite{daghofer.nicholson.12}] the hopping parameters in electron volts are given as: $t_{1} = -0.08$, $t_{2} = 0.1825$, $t_{3} = 0.08375$, $t_{4} = -0.03$, $t_{5} = 0.15$, $t_{6} = 0.15$, $t_{7} = -0.12$, $t_{8} = 0.06$, $t_{9} = 0.0$, $t_{10} = -0.024$, $t_{11} = -0.01$, $t_{12} = 0.0275$ and $\epsilon_{0} = 0.75$. Average number of particles in the system $n = 4$ is attained for $\mu = 0.4748$.

\end{document}